\documentclass[11pt,twoside]{article}

\usepackage{asp2006}
\usepackage{epsf}
\usepackage{psfig}
\usepackage{lscape}
\usepackage{graphicx}

\begin{document}
\title{High Resolution Spectropolarimetry of the H$_\alpha$ Line: Obscured Stars and Absorptive Polarization}   
\author{D.M. Harrington \& J.R. Kuhn}   
\affil{University of Hawaii, Institute for Astronomy, 2680 Woodlawn Drive, Honolulu, HI 96822}    

\begin{abstract} 

     The near-star environment around obscured stars is very dynamic. Many classes of stars show evidence for winds, disks, inflows and outflows with many phenomena occurring simultaneously. These processes are involved in stellar evolution, star and planet formation, and influence the formation and habitability of planets around host stars. Even for the nearest stars, this region will not be imaged even after the completion of the next generation of telescopes. Other methods for measuring the physical properties of circumstellar material must be developed. The polarization of light across spectral lines is a signature that contains information about the circumstellar material on these small spatial scales. We used the HiVIS (R=13000 to 50000) and ESPaDOnS (R=68000) spectropolarimeters to monitor several classes of stars on over a hundred nights of observing from 2004-2008. In 10/30 classical Be stars, the traditional broad depolarization morphology is reproduced, but with some additional absorptive effects in 4 of these 10 stars. In Herbig Ae/Be stars roughly 2/3 of the stars (14/20) with strong absorptive components (either central or blue-shifted) showed clear spectropolarimetric signatures typically centered on absorptive components of the spectral lines. They were typically 0.3\% to 2\% with some signatures being variable in time. Post-AGB and RV-Tau type evolved stars showed very strong absorptive polarimetric effects (5/6 PAGB and 4/4 RVTau) very similar to the Herbig Ae/Be stars. These observations were inconsistent with the traditional scattering models and inspired the development of a new explanation of the observed polarization. This new model, based on optical-pumping, has the potential to provide direct measurements of the circumstellar gas properties.  

\end{abstract}



\section{Introduction}

	The formation and evolution of stars is a dynamic process involving circumstellar matter. Over the course of a few million years, the circumstellar gas and dust around a young star will either accrete onto the star, turn in to planets, or dissipate in the form of winds and jets. There is evidence of simultaneous accretion, outflows, ionized disk structures, and strong stellar winds in young stars. All of these processes influence the environment of the planets around these stars. Similarly, in evolved or rapidly rotating stars, the outer layers are shed to form disks, winds or envelopes. There are only a few techniques that can put meaningful constraints on the environment very close to a star. Spectropolarimetry is such a technique but it is still considered a difficult specialty field. There are only two high-resolution instruments (R$>$10000), ESPaDOnS and HiVIS on large telescopes (over 3m) capable of these measurements (Donati et al. 1997, Harrington \& Kuhn 2008a). However, this technique is a powerful probe of small spatial scales, being sensitive to the geometry and density of the material very close the central star. In a typical model, material in the region a few to several stellar radii away from the star produces the polarization. Even for the closest young stars, these spatial scales are smaller than 0.1 milliarcseconds across and will not be imaged directly, even by the next generation telescopes. Since the circumstellar material is involved in accretion, outflows, winds and disks, with many of these phenomena happening simultaneously, spectropolarimetry can put unique constraints on the types of densities and geometries of the material involved in these processes.  

\begin{figure}
\begin{center}
\includegraphics[width=0.23\linewidth, angle=90]{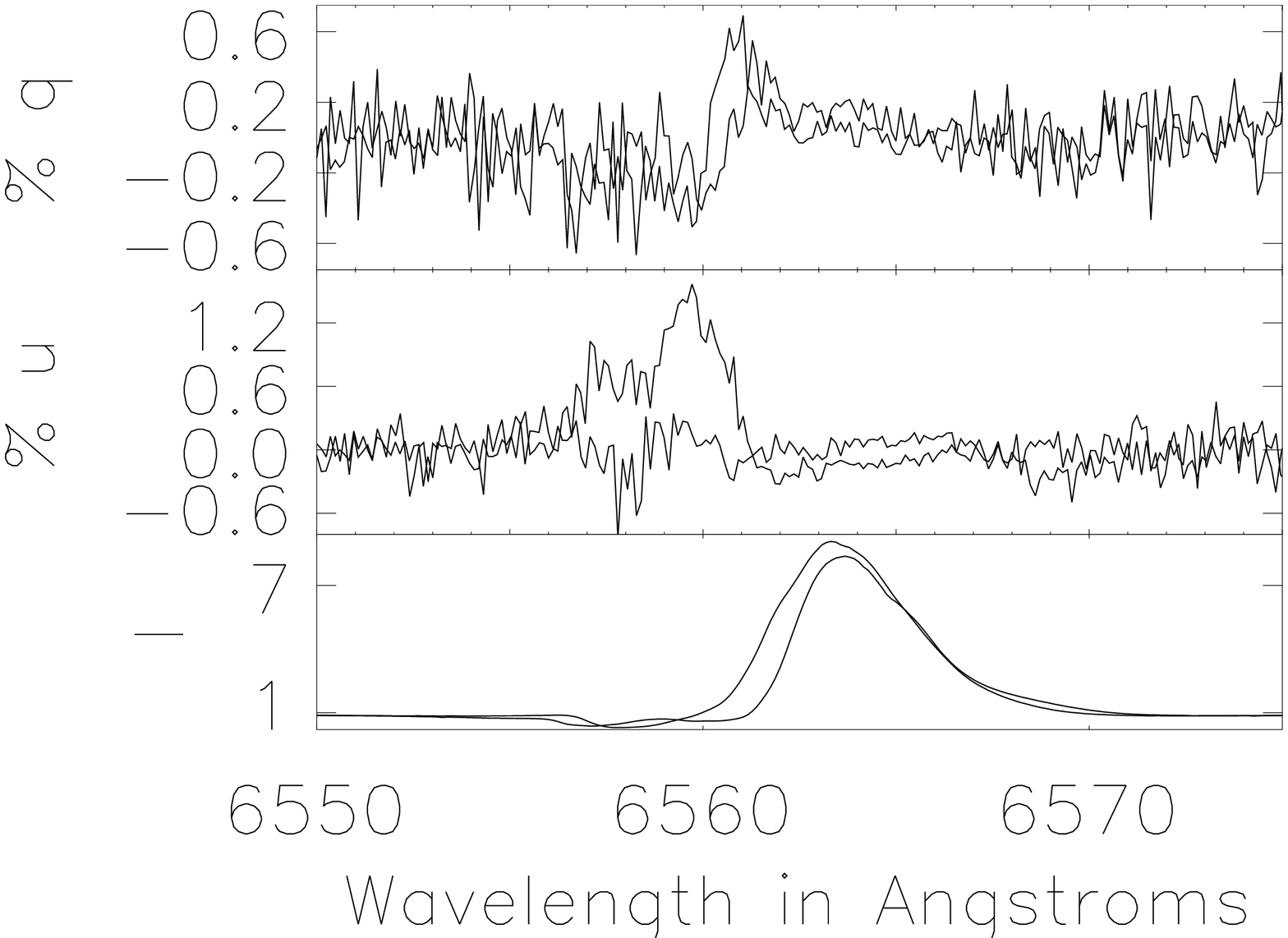}  
\includegraphics[width=0.23\linewidth, angle=90]{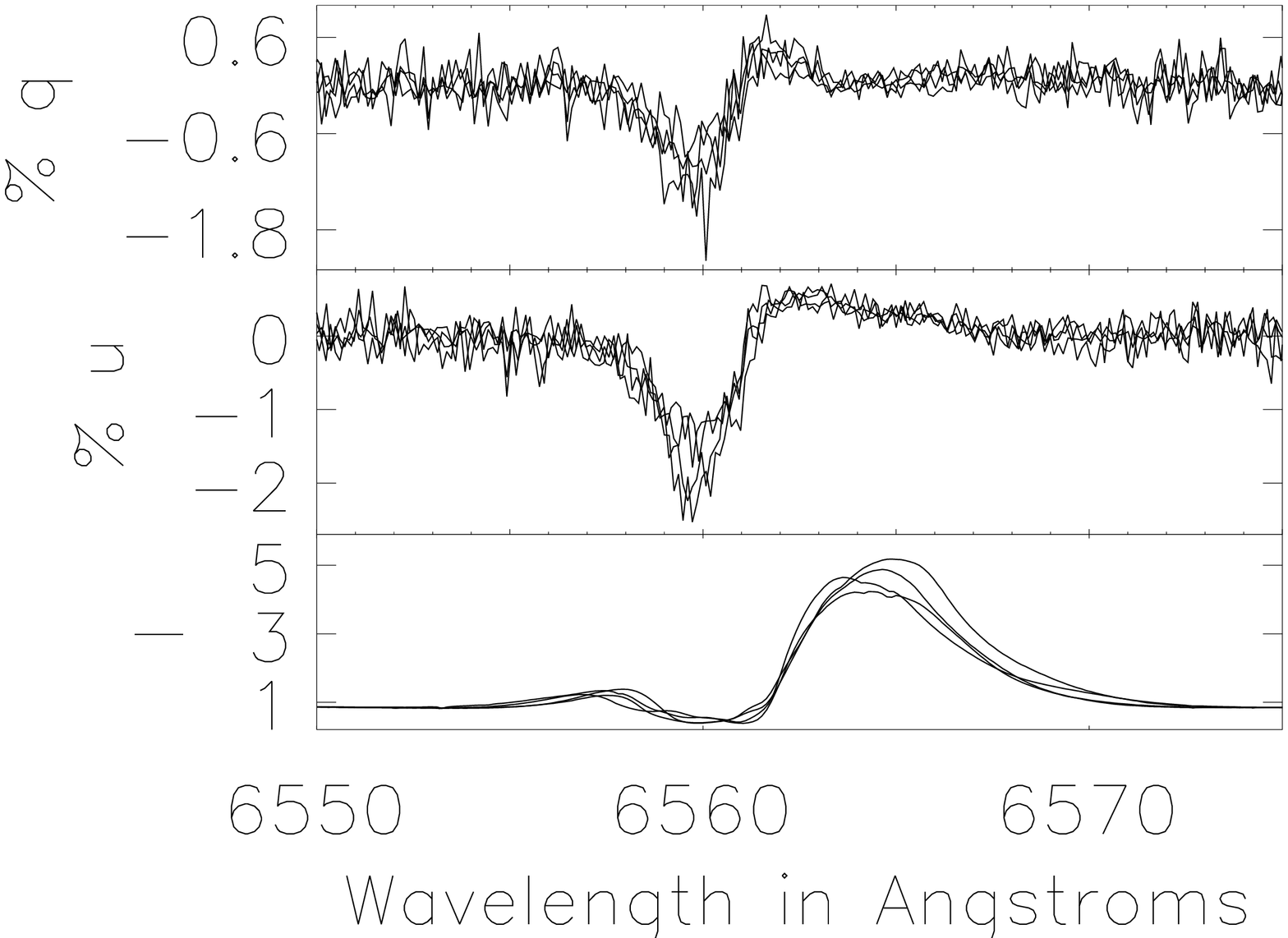}  
\includegraphics[width=0.23\linewidth, angle=90]{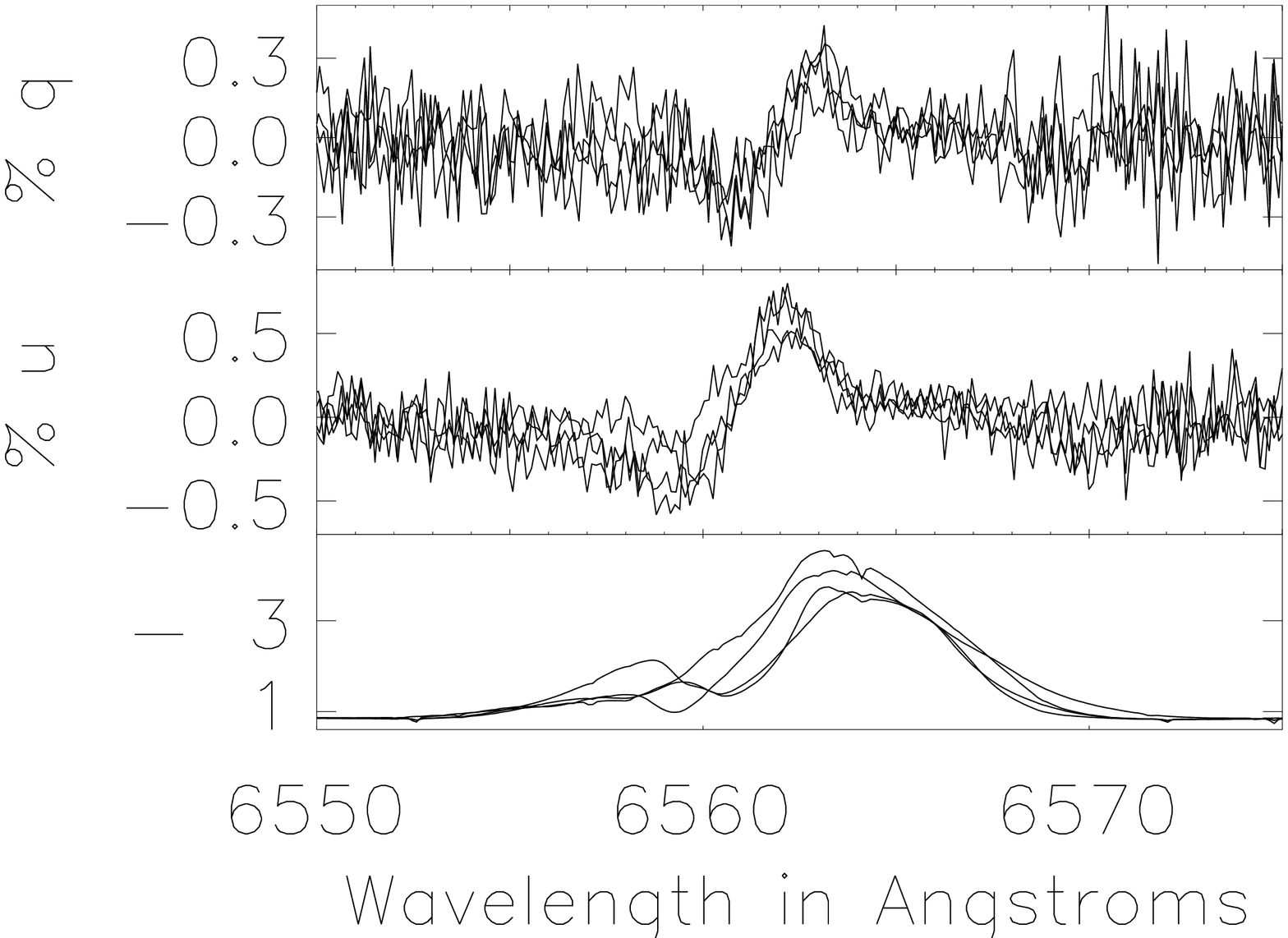}  \\
\includegraphics[width=0.23\linewidth, angle=90]{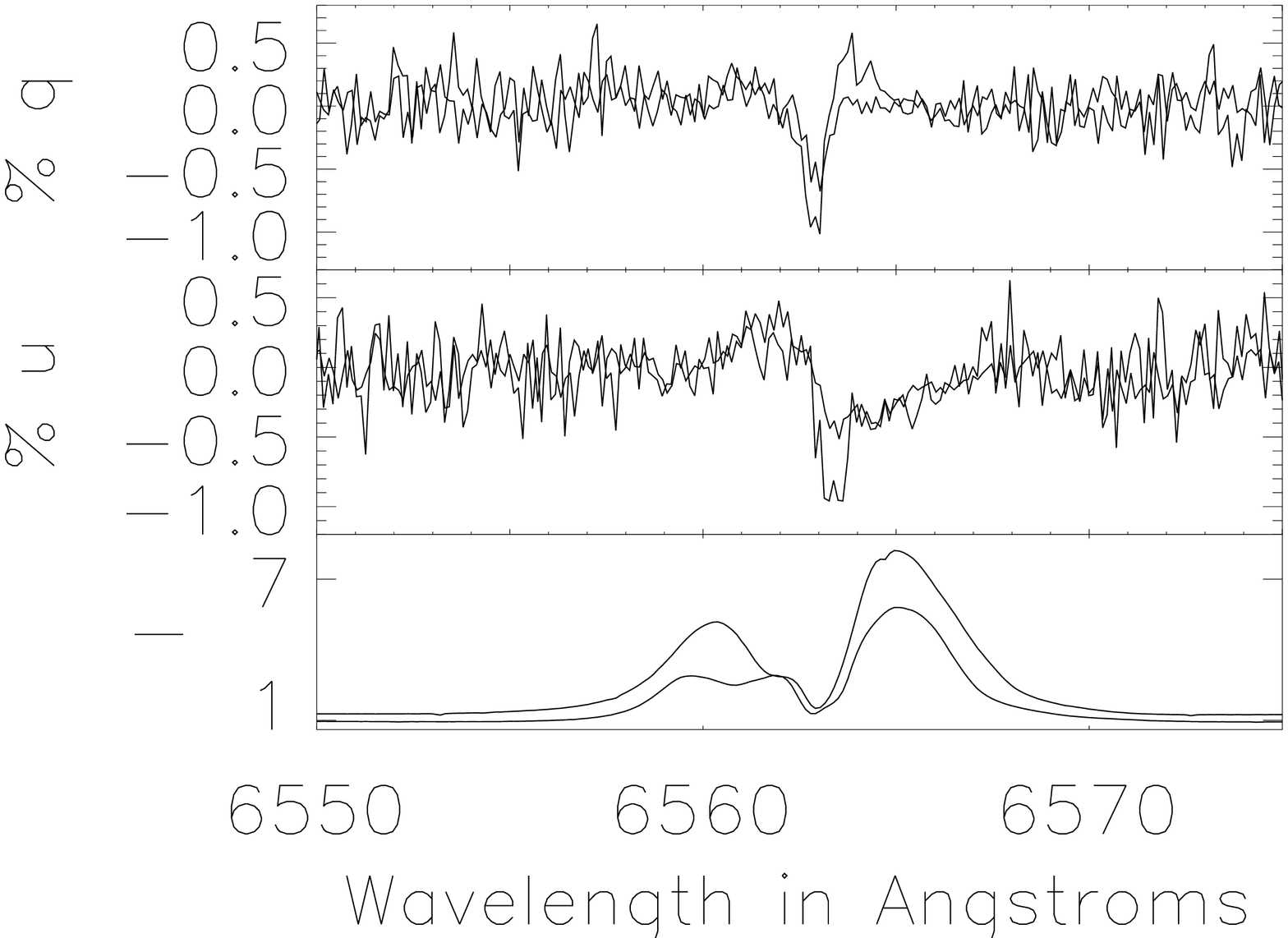} 
\includegraphics[width=0.23\linewidth, angle=90]{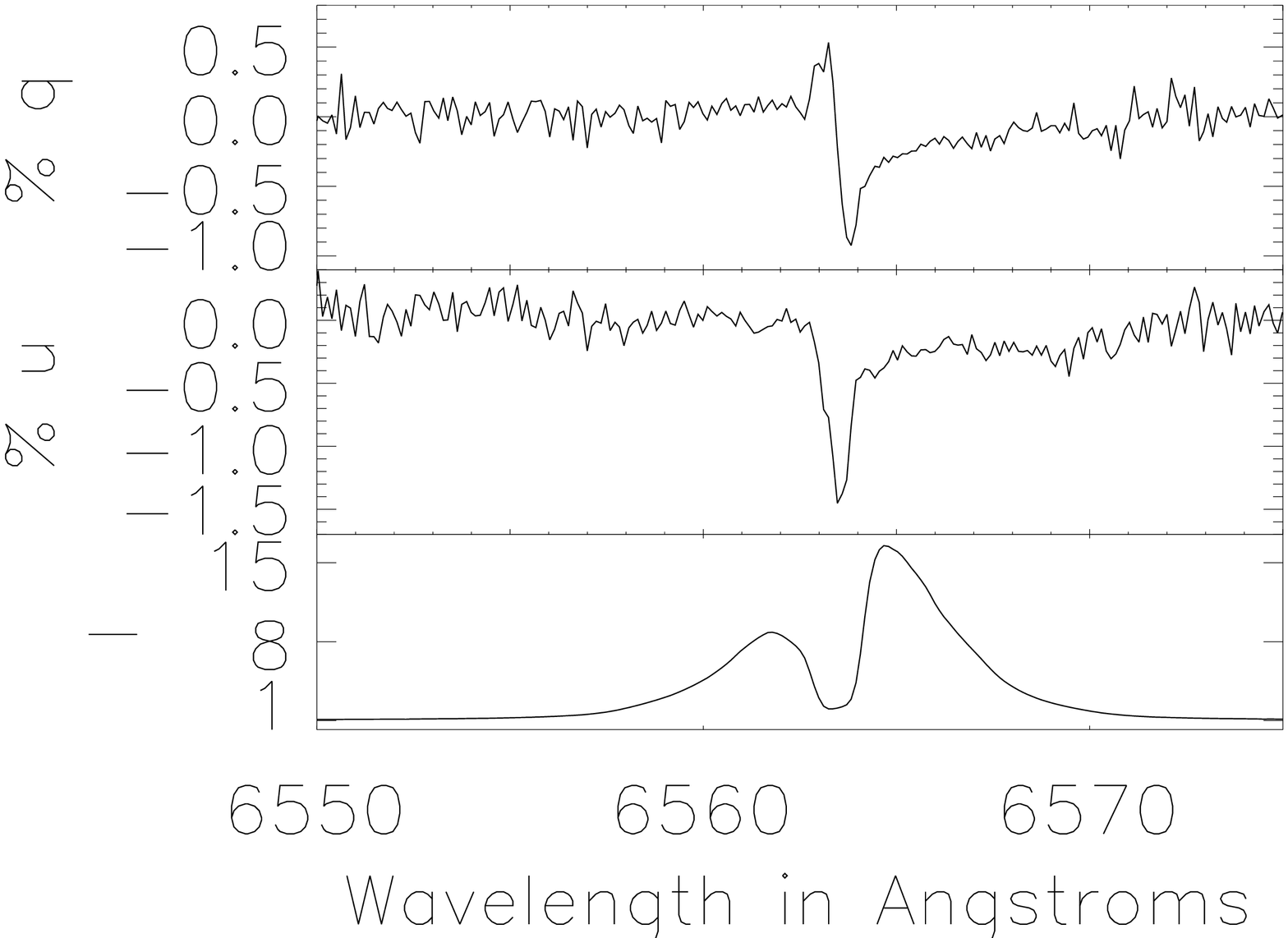}  
\includegraphics[width=0.23\linewidth, angle=90]{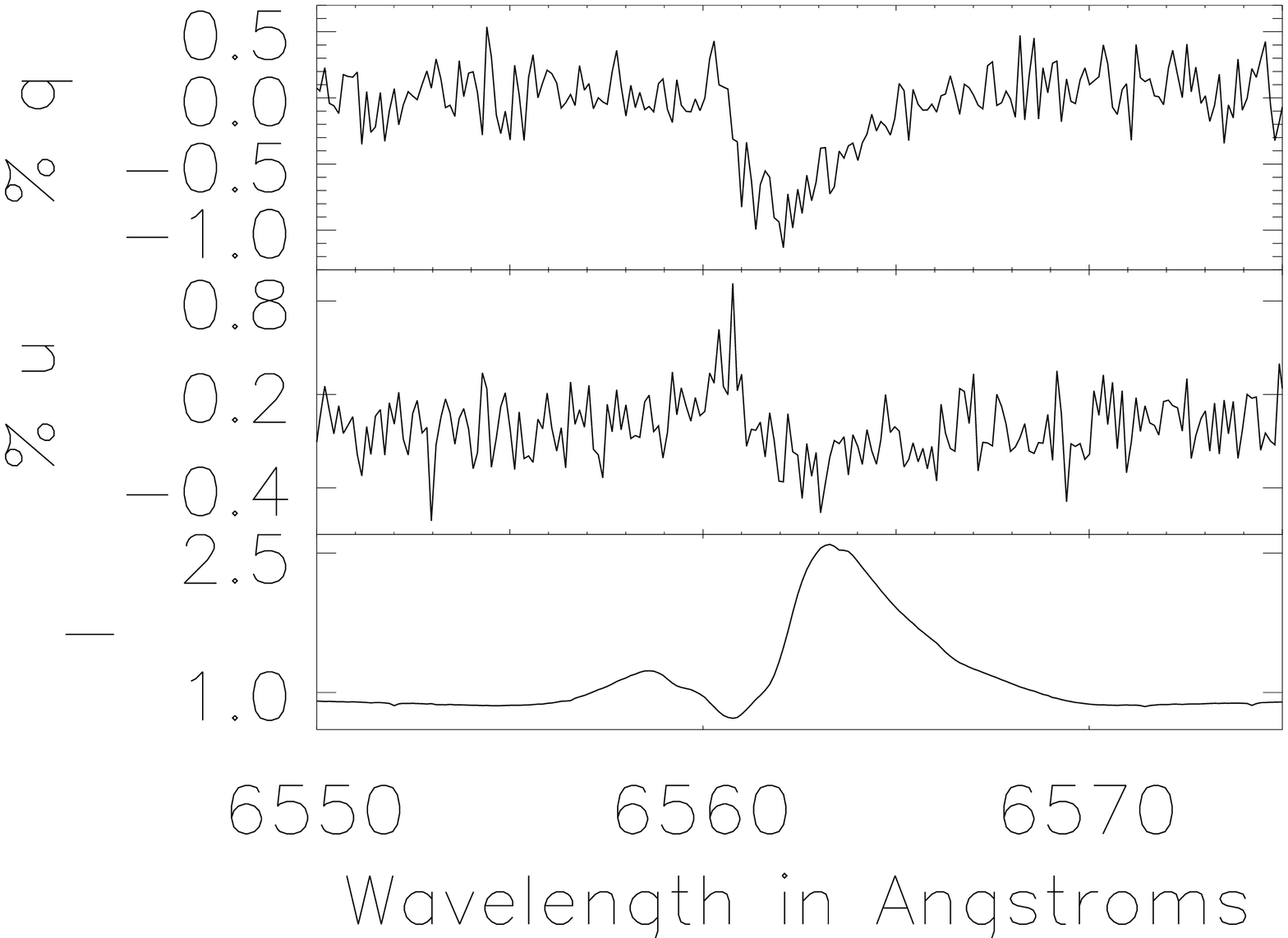} \\
\includegraphics[width=0.23\linewidth, angle=90]{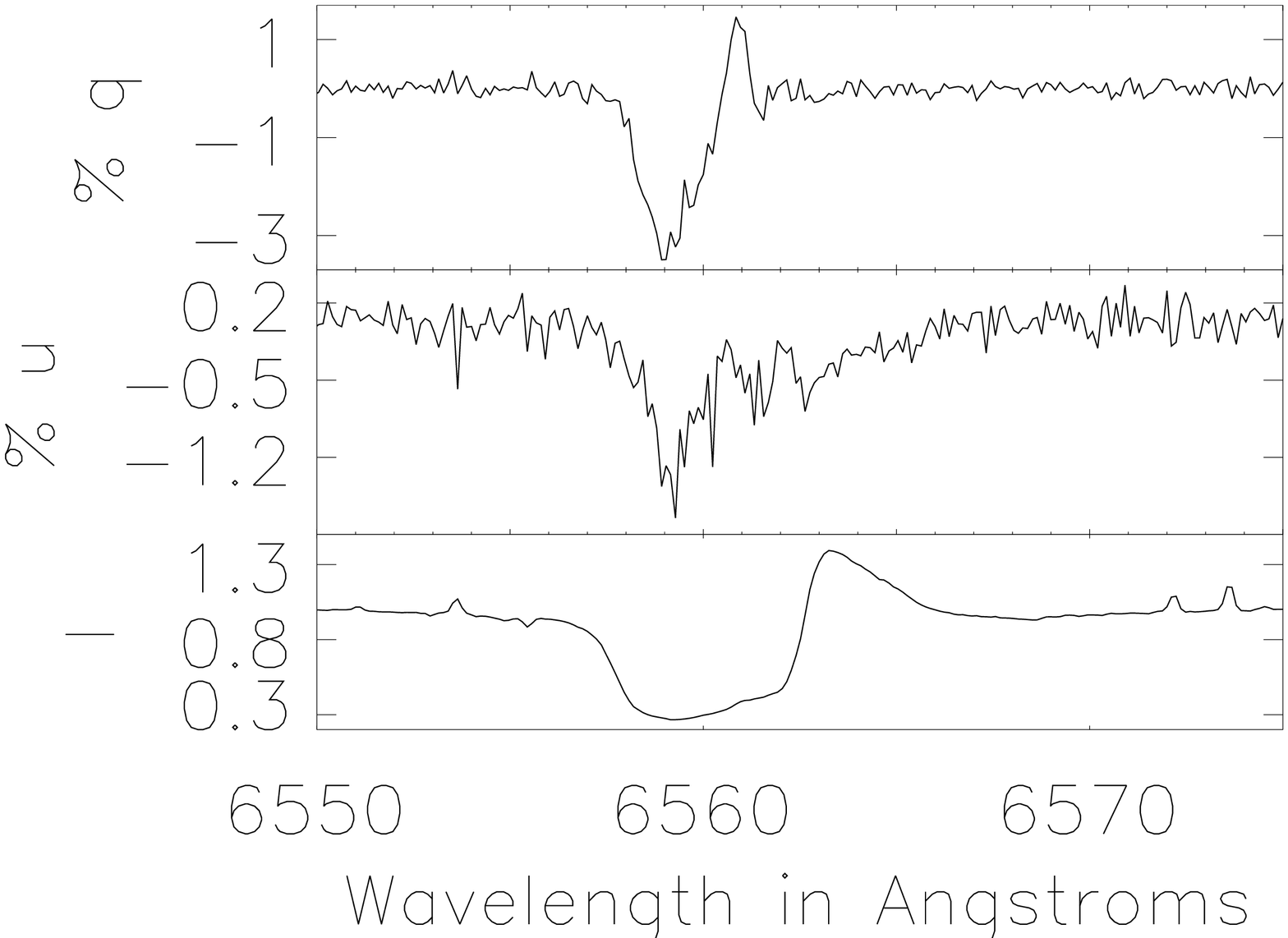} 
\includegraphics[width=0.23\linewidth, angle=90]{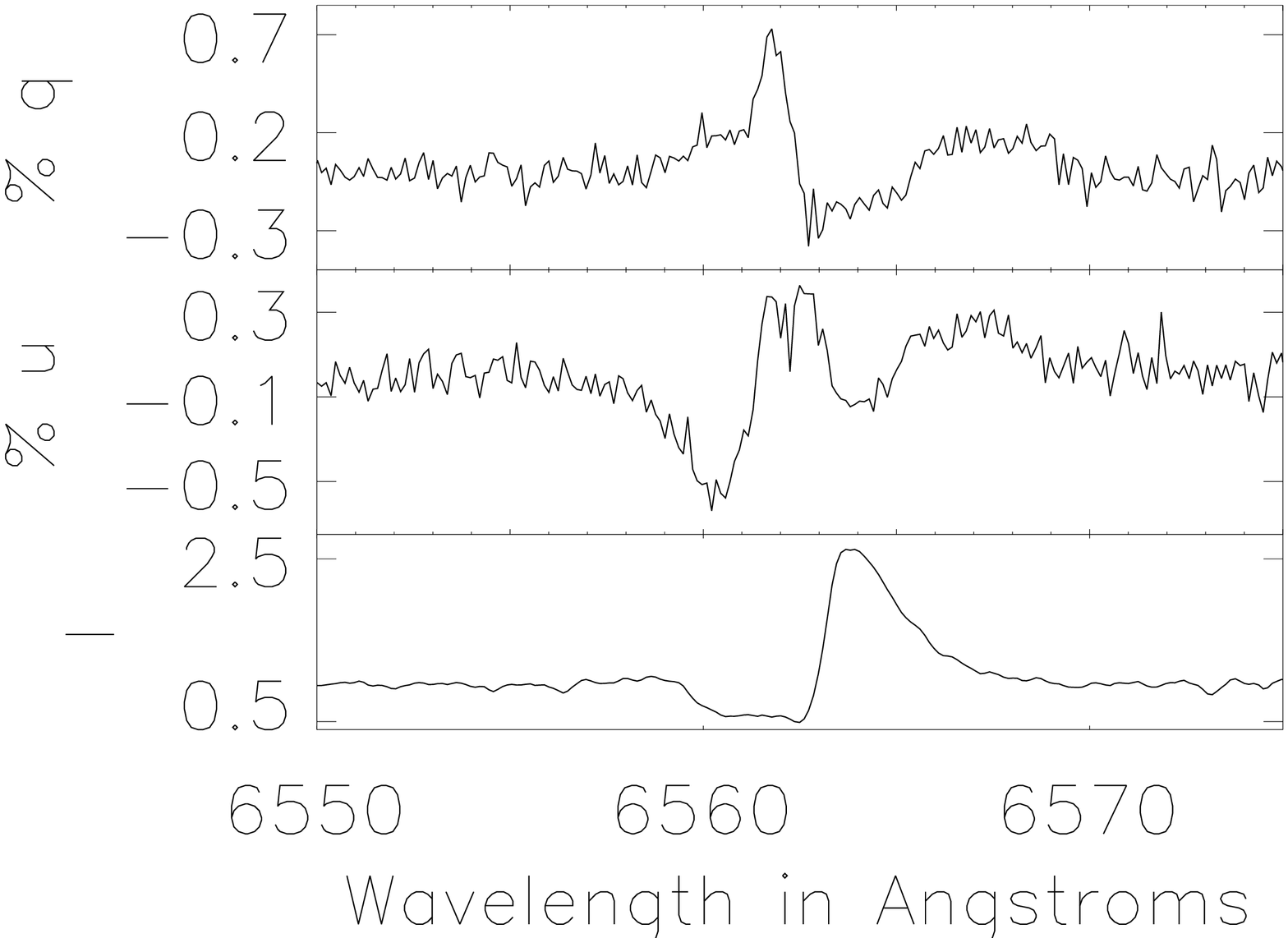}  
\includegraphics[width=0.23\linewidth, angle=90]{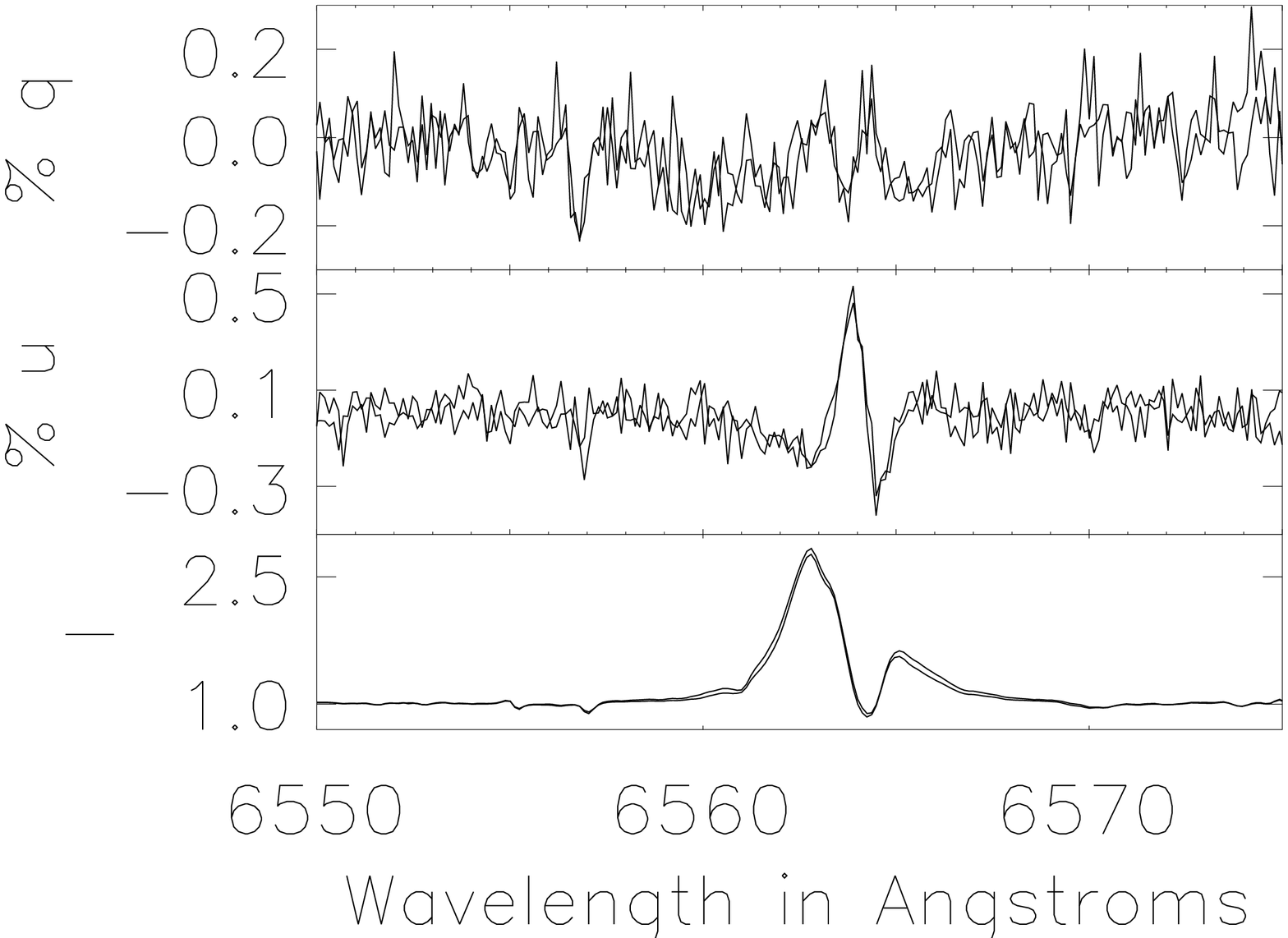} \\
\includegraphics[width=0.23\linewidth, angle=90]{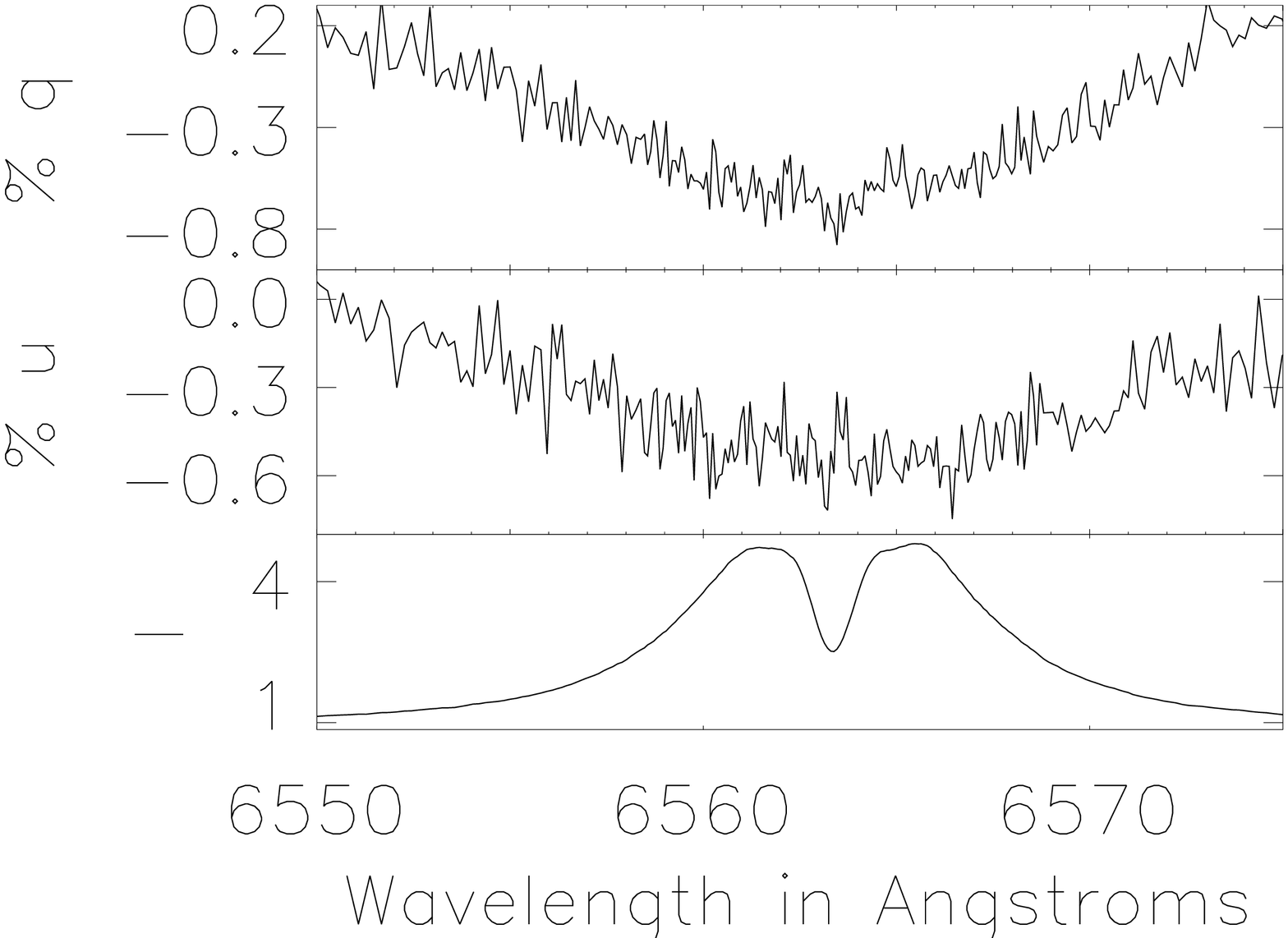} 
\includegraphics[width=0.23\linewidth, angle=90]{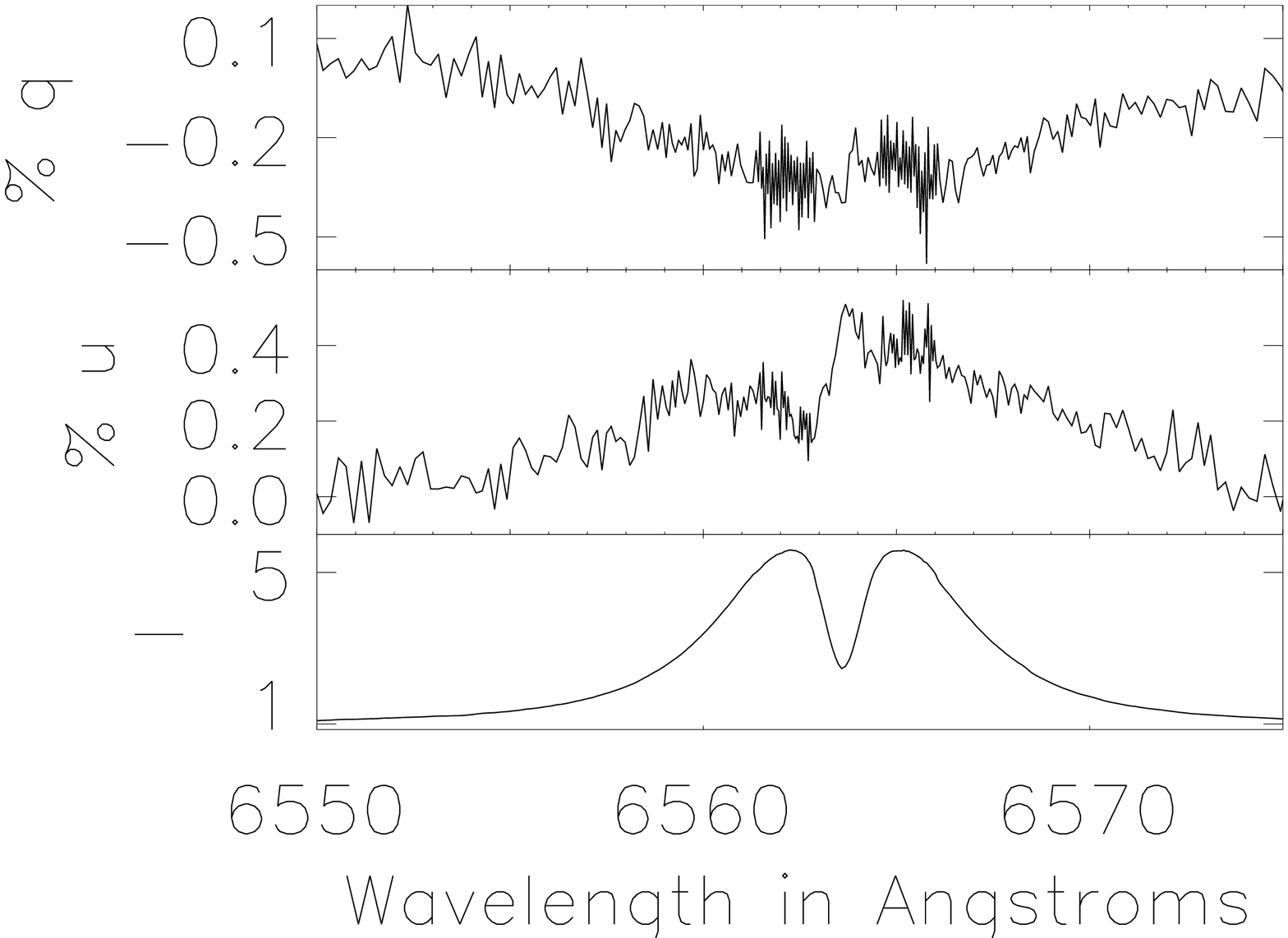}  
\includegraphics[width=0.23\linewidth, angle=90]{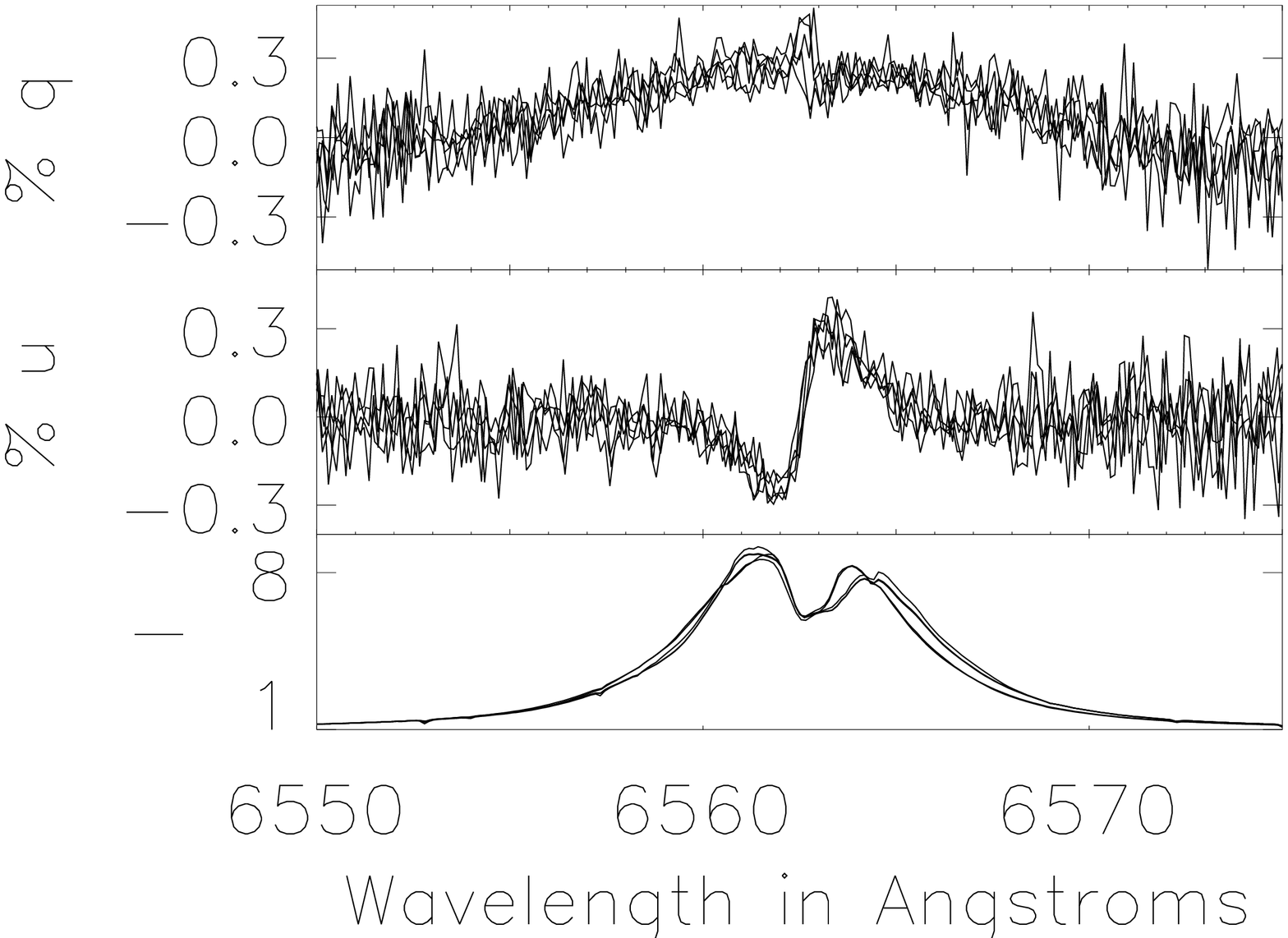} \\
\caption{This figure shows ESPaDOnS spectropolarimetry of Herbig Ae/Be stars, Post-AGB, RV-Tau and Be type stars. The first two rows show the Herbig Ae/Be stars AB Aurigae, MWC 480, HD 163296, MWC 120, MWC 158 and HD 144432 from left to right. The third row shows the Post-AGB/RV-Tau type stars 89 Her, SS Lep and U Mon. The last row shows the Be stars MWC 143, $\psi$ Per and the Herbig Be star MWC 361 for comparison. Very distinct morphologies are seen. The Herbig stars showed nearly ubiquitous polarization-in-absorption (detected in roughly 2/3 of the stars with strong absorptive components and none of the Herbig stars with other line profile types). In the Post-AGB \& RV-Tau stars absorptive effects were ubiquitous (9/10). The Be stars in the last row illustrate the broad ``depolarization'' morphology but with modifications. In our survey 4/10 of these detections showed additional absorptive effects not seen at lower spectral resolution. These observations represent an order-of-magnitude gain in spectral resolution at high signal-to-noise as no flux-dependent wavelength averaging is required to achieve the needed sensitivity. This data set shows the ubiquity of absorptive polarization effects.}
\end{center}
\end{figure}

	With high spectral resolution ($\frac{\lambda}{\delta\lambda}>$10000) clear changes across individual emissive or absorptive line components are observable. Circumstellar disks, rotationally distorted winds, magnetic fields, or asymmetric radiation fields (optical pumping) are all used to explain these signatures. In most models, the polarization change comes directly from the circumstellar material and can be used as a proxy for the physical properties of that material. Spectropolarimetric signals measured to date are small, typically 0.1\% to 1\%. Measuring these signals requires very high signal to noise and careful control of systematics. 
		
	There are many models of linear spectropolarimetric effects. Analytical studies showed spectropolarimetric effects from circumstellar disk scattering (McLean 1979, Wood et al. 1993). Recent modelling of absorptive and scattering processes has shown a wealth of spectropolarimetric effects from disks, winds, and envelopes (cf. Harries 2000, Ignace 2004, Vink et al. 2005, Kuhn et al. 2007). The depolarization effect is caused by unpolarized line emission from stellar envelopes diluting a polarized stellar continuum (McLean 1979). Small clumps in a stellar wind scatter and polarize significant amounts of light can enhance the polarization at that clump's specific velocity and orientation (Harries 2000). Magnetic fields create polarized line profiles through the Zeeman effect. Thin disks can differentially scatter light and produce broad spectropolarimetric profiles (Wood et al. 1993, Wood Brown \& Fox 1994). Optically pumped gas can produce polarization through absorption (Kuhn et al. 2007).

\section{Observations}

	We have recently built a high-resolution polarimeter for the HiVIS spectrograph (R 10,000-50,000) on the 3.67m AEOS telescope on Haleakala, HI (Harrington et al. 2006, Harrington \& Kuhn 2008a). This was used on roughly one hundred nights from 2004-2008 for a large spectropolarimetric survey. In addition to these observations, several nights equivalent of observations as well as archival observations using the ESPaDOnS spectropolarimeter (R=68,000) on the 3.6m Canada-France-Hawaii telescope were used to supplement the survey. The initial results focused on the H$_\alpha$ line in several classes of star. Initially, well-known bright stars for close study but the target list was expanded to include many stellar types including Classical Be, Herbig Ae/Be, Post-AGB, RV-Tau and other bright emission-line stars. The H$_\alpha$ line in these stars is typically an emission line with additional absorptive components. Stars with P-Cygni profiles showed strong variability of the absorptive component of the line.

	The main observational results of the survey have been presented in Harrington 2008 and Harrington \& Kuhn 2007, 2008b and c. In 10/30 Be type stars, the traditional broad depolarization morphology was found. However, at high spectral resolution some additional polarimetric effects were seen in the absorptive components of 4 of these 10 stars. In the Herbig Ae/Be stars roughly 2/3 of the stars with strong absorptive components (14/20 with either central or blue-shifted absorption) showed clear spectropolarimetric signatures. The detected effects were centered on absorptive components of the line and was typically confined to the wavelengths corresponding to strong absorption. The magnitude was typically 0.3\% to 2\% with some signatures being variable in time. Post-AGB and RV-Tau type evolved stars also showed ubiquitous strong absorptive polarimetric effects (5/6 PAGB and 4/4 RVTau) very similar to the Herbig Ae/Be stars. Figure 1 shows examples of the absorptive polarization effects in Herbig Ae/Be stars, Post-AGB stars and classical Be stars. These observations were inconsistent with the traditional scattering models. Both the disk scattering and depolarization models produce broad spectropolarimetric profiles, not profiles confined to absorptive components. This inspired the development of a new explanation of the observed polarization (Kuhn et al. 2007) based on optical-pumping.

\begin{figure}
\begin{center}
\includegraphics[height=0.45\linewidth, angle=90]{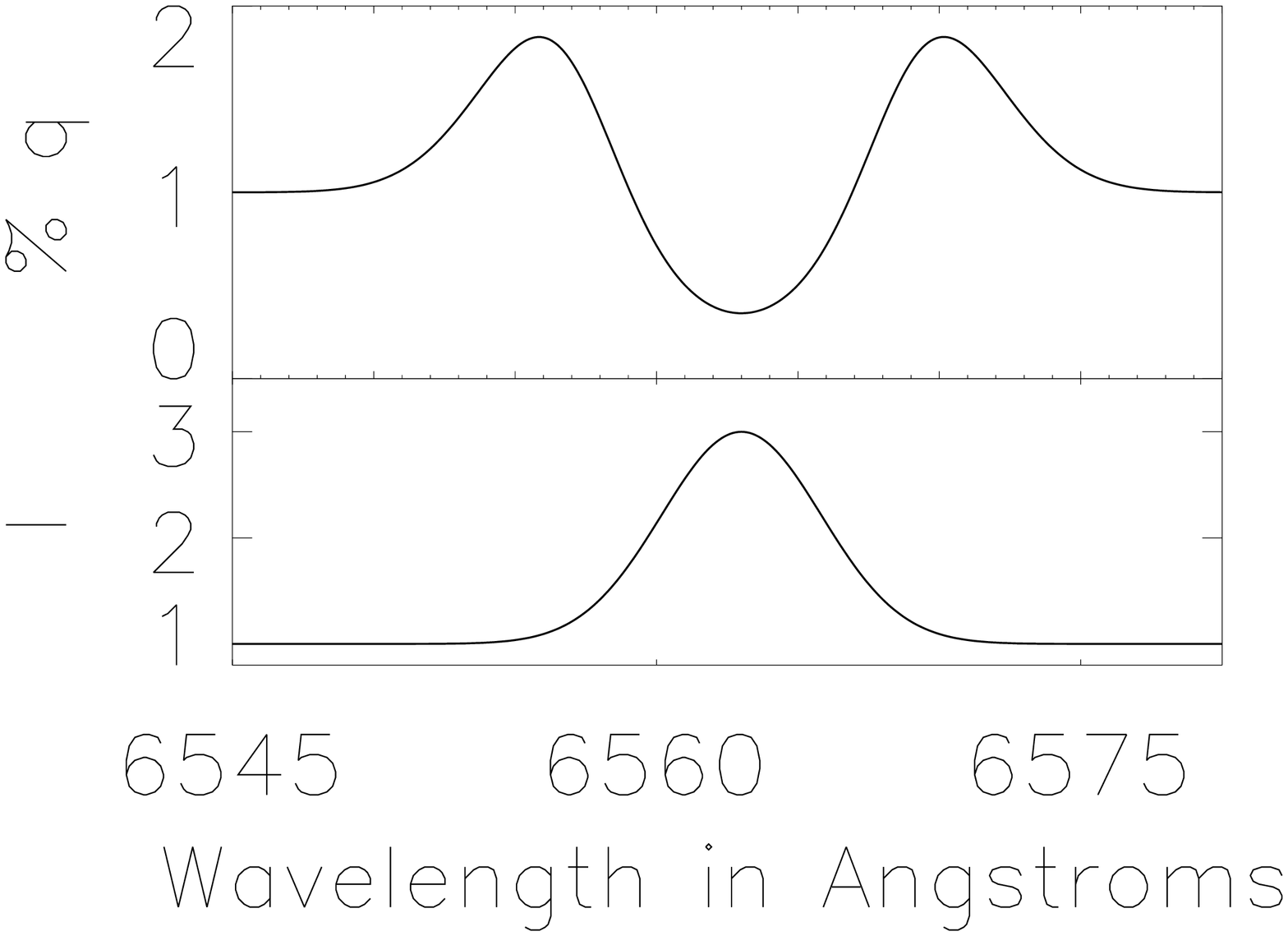}  
\includegraphics[width=0.45\linewidth, angle=0]{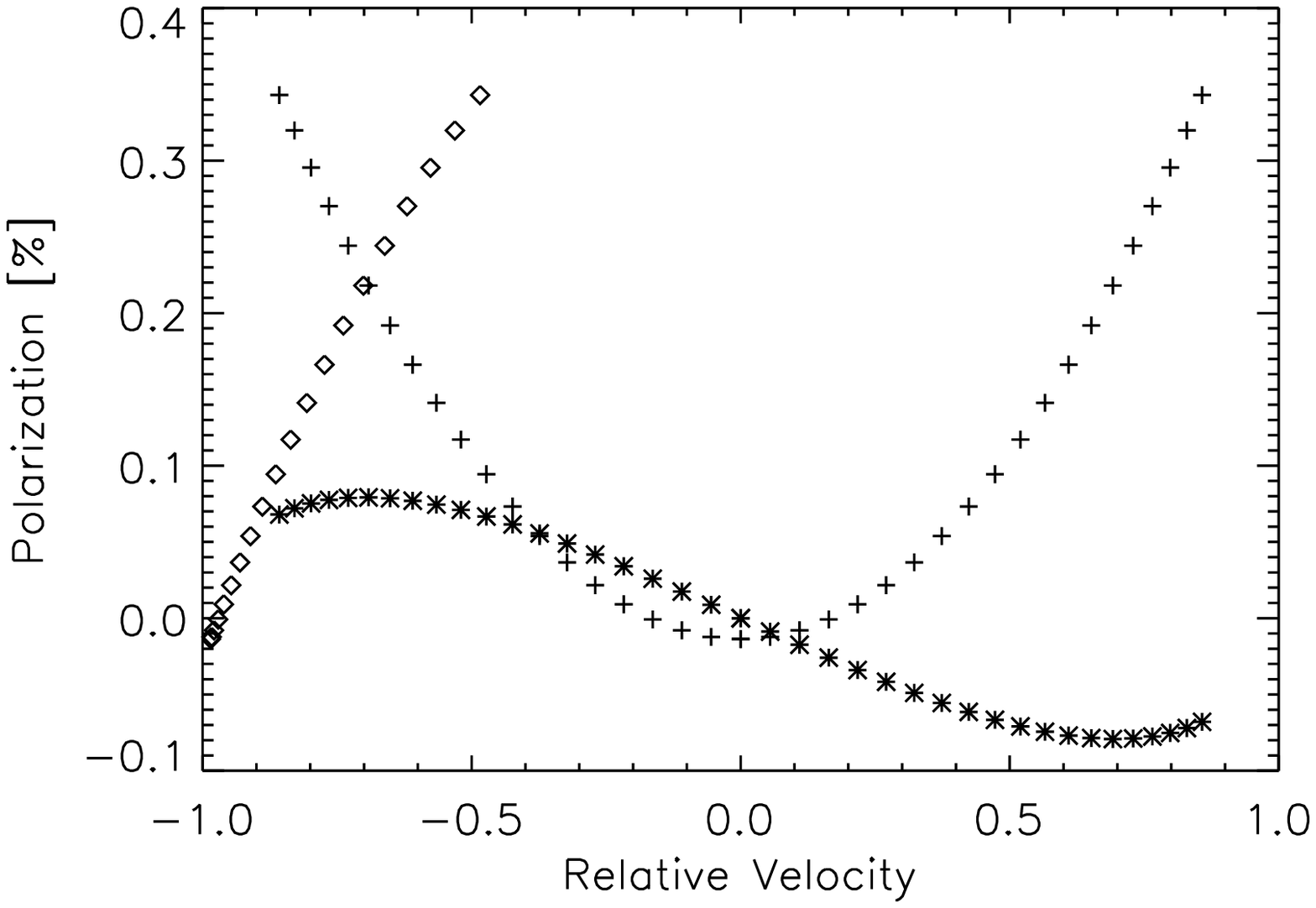}  \\
\caption{This figure illustrates the disk-scattering and optical-pumping theories used to explain spectropolarimetric line profiles. The disk-scattering model produces spectropolarimetric profiles that are broader than the emission line. The models produce symmetric, double-peaked profiles for a simple keplerian disk. Since the observations of Herbig Ae/Be and Post-AGB stars show the polarization being almost entirely contained inside the absorptive component and varying with the absorption, a new model based on optical pumping was developed. In this model, the absorption produces the polarization and the spectropolarimetric profiles are determined by the velocity and geometry of the absorbing gas. The right panel, reproduced with the kind permission of ApJL, illustrates this. See Kuhn et al. 2007 for details.}
\end{center}
\end{figure}

\section{A New Model - Optical Pumping}

	The disk-scattering and depolarization models predict signatures centered on the H$_\alpha$ line. The disk-scattering signature must broader than the emission as the effect comes from the doppler-redistribution of the emission line (Wood et al. 1993, Vink et al. 2005). This is illustrated in the left panel of figure 2. The depolarization model has the emission dilute a polarized continuum producing an effect at least as wide as the emission. In almost all of our stars, the effect occurred in and around the absorptive component with the morphology being dominated by the polarization in the absorptive component. The polarization of the emission peak and of the line wings was nearly identical to the continuum polarization in most stars. This problem led us to explore an alternative explanations that has the absorbing material also be the polarizing material. Optical pumping is a very robust phenomena used in the lab for decades that produces polarized absorption in strongly irradiated environments (Happer 1972). We developed a new model where the anisotropic stellar radiation causes the absorbing material to polarize the transmitted light (Kuhn et al. 2007). The key difference between this optical pumping model and the scattering model is that only the absorbing material is responsible for the changing polarization, whereas the scattering models require scattered light from the entire circumstellar region producing polarization effects across the entire line. An example of an optical pumping calculation is shown in the right panel of figure 2 from Kuhn et al. 2007. The calculation shows the polarization for an orbiting or outflowing cloud obscuring the photosphere of a star as a function of it's projected radial velocity. This new model is simple, deterministic and produces polarization in absorption as we have observed in several classes of star.

\acknowledgements 

	This program was supported by the NSF AST-0123390 grant, the University of Hawaii and the AirForce Research Labs. Some of this research used the facilities of the Canadian Astronomy Data Centre operated by the National Research Council of Canada with the support of the Canadian Space Agency. This program also made use of observations obtained at the Canada-France-Hawaii Telescope (CFHT) which is operated by the National Research Council of Canada, the Institut National des Sciences de l'Univers of the Centre National de la Recherche Scientifique of France, and the University of Hawaii. These CFHT observations were reduced with the dedicated software package Libre-Esprit made available by J. -F. Donati. The Simbad data base operated by CDS, Strasbourg, France was very useful for this survey.



\begin{thebibliography}{}
\bibitem{don99}   Donati J.F. et al., 1997, MNRAS, 291, 658
\bibitem{har00}     Harries T.J., 2000, MNRAS, 315, 722
\bibitem{hap72}  Happer W., 1972, Rev. Mod. Phys., 44, 169
\bibitem{hk06}    Harrington D.M. et al., 2006 PASP, 118, 845
\bibitem{hk07}   Harrington D.M. \& Kuhn J.R., 2007, ApJL, 667, L89
\bibitem{har08}  Harrington D.M., 2008, Univ. Hawaii PhD Thesis
\bibitem{har08a}  Harrington D.M. \& Kuhn J.R., 2008, PASP, 120, 89
\bibitem{hk08b}  Harrington D.M. \& Kuhn J.R., 2008 ApJS, Submitted
\bibitem{hk08c}  Harrington D.M. \& Kuhn J.R., 2008 ApJL, Submitted
\bibitem{ign04}    Ignace R. et al., 2004, ApJ, 609, 1018
\bibitem{kuh07}   Kuhn J.R. et al., 2007, ApJL, 668, L63
\bibitem{mcl79}    McLean I.S., 1979, MNRAS, 186, 265
\bibitem{vin05}    Vink J.S., 2005, A\&A, 430, 213
\bibitem{wo93}     Wood K. et al., 1993, A\&A, 271, 492
\bibitem{wo94}     Wood K. \& Brown J.C., 1994, A\&A, 291, 202
\end{thebibliography}
\end{document}